\documentstyle[preprint,tighten,prc,aps,epsfig]{revtex}

\newcommand{\be}{\begin{equation}}
\newcommand{\ee}{\end{equation}}
\newcommand{\bfeps}{\mbox {\boldmath $\epsilon$}}
\newcommand{\bfp}{\mbox {\boldmath $p$}}
\newcommand{\bfk}{\mbox {\boldmath $k$}}
\newcommand{\bfs}{\mbox {\boldmath $S$}}
\newcommand{\bfg}{\mbox {\boldmath $G$}}
\newcommand{\bfzeta}{\mbox {\boldmath $\zeta$}}
\newcommand{\bfsigma}{\mbox {\boldmath $\sigma$}}

\newcommand{\mix}{$a^0_0\,$$-$$f_0$\, }
\newcommand{\pieta}{$\pi^0$$-$$\eta$}
\newcommand{\aofomix}{\mbox {\boldmath $a_0\,$-$f_0$}}

\newcommand{\vpn}{\mbox {\boldmath $\vec p n\to d a_0$}}

\begin{document}
\preprint{FZJ-IKP-TH-2002-05} 
\title{Aspects of \aofomix\ mixing in the reaction \vpn\ }

\author{A.E. Kudryavtsev$^1$, V.E. Tarasov$^1$,
J. Haidenbauer$^2$, C. Hanhart$^2$, and J. Speth$^2$}
\address{$^1$Institute of Theoretical and Experimental Physics, \\
B. Cheremushkinskaya 25, 117259 Moscow, Russia \\
$^2$Institut f\"ur Kernphysik, Forschungszentrum J\"ulich,
D-52425 J\"ulich } 

\date{\today}
\maketitle

\begin{abstract}
Some aspects of \mix mixing effects in the reaction
${\vec p} n\to da^0_0$ with perpendicular polarized proton beam are
discussed. An angular--asymmetry parameter $A$ is defined to study those
effect. It is shown that, for energies close to the production threshold, 
the angular--asymmetry parameter $A(\theta, \varphi)$ is proportional to
the \mix mixing amplitude for arbitrary polar and azimuthal angles
$\theta$ and $\varphi$ of the outgoing $a_0$ meson.
This statement is also valid for arbitrary
energies, but then only at polar angles $\theta=0^0$ and $\theta=90^0$.
The mass dependence of the differential cross section $d\sigma/dm_{\pi^0\eta}$
in the reaction $pn\to d\pi^0\eta$ in the presence of \mix mixing
is also discussed.
\end{abstract}

\pacs{13.60.Le; 13.75.-n; 14.40.Cs}

\section{Introduction}
The nature of the lightest, virtually mass-degenerate, scalar mesons
$a_0$ (980) ($I^G J^{PC}=1^- 0^{++}$) and $f_0$ (980) ($0^+ 0^{++}$)
is an important and still unsolved problem of hadron physics.
The quark structure of
these mesons is not well established at present time.
This issue is also closely related to the very interesting problem
of $a_0\,$-$f_0$ mixing. A dynamical mechanism for this mixing
close to $K\bar K$ threshold was suggested around 20 years ago in
Ref.~\cite{Acha}.
Since that time a number of papers have been published in which different
aspects of $a_0\,$-$f_0$ mixing and the possibilities to measure
this effect have been discussed,
see, for example, Refs.~\cite{Barnes,Acha1,Krepl,Kerb,Kud,Grish}.
Recently, in Ref.~\cite{Close}, some new arguments were presented
in support of a fairly large \mix-mixing intensity.
Analyzing the experimental
data~\cite{Barb,Kirk} on the exclusive production of the $a_0$(980),
$a_2$(1320) and $f_0$(980) resonances in $pp$ collisions at
$\sqrt{s}=29.1$~GeV, the authors of Ref.~\cite{Close} came to the
conclusion that $80\pm 25$~\% of all $a^0_0$ mesons come from the
$f_0$(980).

The results of Ref.~\cite{Close} provide certainly a
fairly strong indication for a large \mix\ mixing. However, for more
solid conclusions, in particular about the quantitative
value of the mixing, a thorough analysis is needed.
In this context it is important to keep in mind that the mixing
can manifest itself in many different physical processes and observables.
Thus, in this paper we want to discuss another
consequence of the \mix mixing effect, namely the angular
asymmetry of the reaction
\be
p\,n \to d\, a^0_0/f_0 \, .
\label{1}
\ee
with unpolarized and polarized proton beam.
Furthermore we study the influence of the finite width of the
scalar mesons on the unpolarized as well as the polarized differential
cross section. All these investigations are performed with special emphasis
on how to extract informations on the $a_0/f_0$ mixing.

In the present paper we will address the $a_0$ and $f_0$ mesons
as resonances. However,
it should be stressed that all the results derived in this paper
do not need any assumption about the nature of those scalar mesons.
Even if these scalar mesons are dynamically generated
\cite{Jans,Oller97}, in the proximity
of the pole the scattering matrix can still be well approximated
by the propagation of quasiparticles corresponding to elementary
fields. The true nature of the
propagating object is then hidden in the effective parameters
of those elementary fields. This was shown rigorously by Weinberg
for the case where inelastic channels are absent \cite{weinberg}.

The paper is organized as follows. In Section 2 we review some common
properties of the production of unstable particles. We start out
from the case of a single particle and then, in Sect. 3, generalize
the formalism to the two channel case relevant for the present study.
Section 4 contains a discussion of the general structure of the primary
production amplitude for small excess energies.
Section 5 is devoted to the study of different observables.
The main emphasis is put on the $\pi \eta$ final state. Two differential
observables are analyzed in detail, namely the double 
differential cross section $d^2\sigma/d\Omega/dm^2$ as well as the
analyzing power. The former observable was studied recently in a series of
papers, however, without investigating the $m^2$ dependence more thoroughly.
The latter, on the other hand, was not discussed in detail before at all.
We close with a short summary and conclusions.

\section{The production of unstable particles}

Let us start with the general expression for the  cross section of the
reaction $NN \to dX \ ,$ where $X$ denotes the decay products of an
unstable meson.
We start the discussion by assuming the presence of one mesonic resonance
only. The generalization to more than one resonances will be done
in the next section. The main purpose of this section is to
introduce our notation and to remind the reader on the impact of
a finite decay width on observables.

 The reaction cross section in the center-of-mass (CM) system
be can expressed as
\be
d\sigma_X = (2\pi)^4 \, \frac{1}{4|\vec p \,|\sqrt{s}} \,
|{\cal A}_X|^2 d\Phi_{\mu+1}(P;p_d,k_1,...,k_\mu) \ .
\label{dsig}
\ee
Here $\vec p $ denotes the CM momentum of the initial protons and
$P$ the total initial four-momentum. The reaction amplitude ${\cal A}_X$
is given by 
\be
{\cal A}_X = W_X(k_1,...,k_\mu)G(m^2){\cal M} \ .
\label{adef}
\ee
where the primary production amplitude is denoted by ${\cal M}$.
We assume the unstable meson, whose propagation is described
by $G(m^2)$, to decay into the $\mu$ particles of the final state $X$
through the vertex function $W_X$. $m^2$ is the total invariant
mass of the final $\mu$-particles system and is given by $m^2=(\sum k_i)^2$,
with $k_i$, $i=1,...,\mu$ being the four-momenta of the $\mu$ decay particles.
The phase space of the final $\mu$ particles and the final deuteron (with
the four-momentum $p_d$) is defined as
\begin{eqnarray}
\nonumber
d\Phi_{\mu+1}(P;p_d,k_1,...,k_\mu) &=& \delta^4(P-p_d-k)\,
 \frac{d^3p_d}{(2\pi)^3 2E_d}
\prod_i \frac{d^3k_i}{(2\pi)^3 2\omega_i} \\
&=&(2\pi)^3dm^2d\Phi_{2}(P;p_d,k)\, d\Phi_{\mu}(k;k_1,...,k_\mu)
\label{dphi}
\end{eqnarray}
where $E_d$ and $\omega_i$ are the energies of the final deuteron and
the $i$-th decay particle, respectively, and $k = \sum k_i$.
The latter recursive formula
allows to study the propagation and the decay of the unstable particles
independently of the production mechanism itself.

It is convenient to introduce what we would like to call
partial spectral functions $\rho_X$, which is given by
\begin{eqnarray}
\rho_X(m^2) :=
(2\pi )^3\int d\Phi_\mu(k;k_1,...,k_\mu)\,\, |G(m^2)\, W_X|^2 \ .
\label{rho}
\end{eqnarray}
Note that unitarity demands that
\be
{\sum_{X}} \, \rho_X (m^2)
 =  -\frac{1}{\pi}{\rm Im} G(m^2) \ , \ \ \int dm^2 {\sum_{X}} \, \rho_X (m^2)
 = 1 \ ,
\label{rhoeig}
\ee
where the sum runs over all decay channels of the unstable particle
that are open at a certain given value of $m^2$.

Using Eqs. (\ref{dsig}) -- (\ref{rho}) we can write
\be
\frac{d^2\sigma_X}{dm^2d\Omega_k}
=\frac{1}{64\pi^2}\left(\frac{|\vec k \, |}{|\vec p \, |}\right)
\frac{1}{s}|{\cal M}|^2\rho_X(m^2) \ ,
\label{csform}
\ee
where $\vec k$ is the relative momentum of the meson
system $X$ with respect to the deuteron. Naturally, $|\vec k \, |$ is
a function of $m^2$, namely
\be
|\vec k \, |=\frac{\lambda^{1/2}(s,m^2,M_d^2)}{2\sqrt{s}},
\label{momenta}
\ee
where $M_d$ denotes the mass of the deuteron and
the function $\lambda$ is given by
\be
\lambda(x,y,z)=(x-y-z)^2-4yz.
\ee

In this paper we neglect a possible final state interaction (FSI) of the 
produced mesons with the deuteron (this issue was discussed for the reaction 
$pp\to d a^+_0$ in Ref.~\cite{Oset}). 
Then, close to threshold, the energy dependence of the matrix element $\cal M$ 
is determined by the lowest possible orbital angular momentum $l$ of the meson
system $X$ with respect to the deuteron. Specifically, it will 
be given by $k^l$ with $k=|\vec k \, |$ being the corresponding relative
momentum. 

Let us now consider the case that interests us in the present study, namely 
the decay of a scalar resonance or quasiparticle into two pseudoscalar
mesons. For any given interaction of the
scalar mesons with the pseudoscalars, the physical two point functions for the
scalar mesons can then be constructed by solving the Dyson equation,
where the self-energy $\Sigma$ is given by the leading order two pseudoscalar
loop diagrams. The equation is illustrated in Fig.~\ref{dyson}.
The strongest energy dependence of the loops is introduced by the
unitarity cut and its analytical continuation below the threshold.
The remaining piece can be assumed constant and can be absorbed in the
physical mass $M_R$. Under these assumptions the physical propagators of
the scalar mesons can be described by a Flatt\'e form \cite{flatte},
namely by
\begin{equation}
G(m^2) = \frac{1}{m^2-M_R^2 + iM_R {\sum\limits_{X}} \Gamma_X} \ ,
\label{prop}
\end{equation}
where
$$
\Gamma_X=\frac{|W_X|^2 | \vec k_X \, |}{8\pi M_R^2} \, ,
$$ 
with $\vec k_X$ being the CM momentum of the meson system $X$. 
Below the threshold of the production of the final state $X$ the 
analytic continuation of $\Gamma_X$ is to be used.

\section{The mixing amplitude}

We now turn to the case that is relevant for the present paper,
namely when there are two propagating unstable particles with
different isospins. Then the self-energy $\Sigma$ exhibits a matrix
structure in isospin space.
If we now allow the propagating particles to mix then the self-energy
matrix develops non-diagonal elements. Note, since the mixing particles
$a_0$ and $f_0$ are essentially mass degenerate
all the mixing should be completely dominated by its effect on the
meson propagation itself.
Along the lines of this reasoning we
completely neglect any isospin violation in the primary production amplitude
$\cal M$ and in the coupling of the pseudoscalar mesons to the resonances.
 In short, we assume the resonances to be produced in pure isospin
states. The physical states, however, are those that propagate.
The physical states of $a_0$ and $f_0$ are consequently no longer pure
isospin states.

We start from the inverse propagator in the isospin basis,
\begin{eqnarray}
{\mathbf{G^{-1}}} (m^2) = \left(\begin{array}{cc}
(m^2-m_1^2-\Sigma_{11}) & -\Sigma_{10} \\ -\Sigma_{01} & (m^2-m_0^2-\Sigma_{00} )
\end{array} \right)
=: \left( \begin{array}{cc}
g_{11} & g_{12} \\ g_{12} & g_{22}
\end{array} \right) \ ,
\nonumber
\end{eqnarray}
where $m_1$ and $m_0$ denote the bare masses in the isospin 1 and 0
channel, respectively.
The diagonal elements of the self-energy are given by
\begin{eqnarray}
\Sigma_{11} &=& \bar \Pi_{\bar K K}^{1}+\Pi_{\pi\eta} \\
\Sigma_{00} &=& \bar \Pi_{\bar K K}^{0}+\Pi_{\pi\pi}
\end{eqnarray}
and the isospin breaking is generated by
\be
\Sigma_{10} = \Sigma_{01} = \alpha +
\Pi_{K^+ K^-}-\Pi_{\bar K^0 K^0} \ .
\label{smix}
\ee
In these formulas $\Pi_{xy}$ denote loops integrals with particles
$x$ and $y$ propagating. The loops are assumed to be renormalized --
the renormalization constants are either absorbed into the physical
masses of the scalar mesons or into the complex quantity $\alpha$
(see also the discussion of this question in Ref.~\cite{Acha2}).
The loops denoted by $\bar \Pi^I_{\bar K K}$ are to be read as
containing contributions from charged kaons as well as neutral kaons,
coupled to the isospin $I$ as indicated in the superscript. 
It is also implied that the physical masses are used in those
loop integrals. The mass difference between the charged and the 
neutral kaons is the source for the isospin breaking generated by
the loops of Eq. (\ref{smix}). Without this difference the two kaon
loops would cancel each other. 
The complex quantity $\alpha$ contains the effect induced by
$\pi$-$\eta$ mixing \cite{Tipp}, as for instance discussed in Ref.~\cite{Kud},
as well as a possibly existing direct $a_0$--$f_0$ transition.
Naturally, the imaginary part of $\alpha$, induced by the \pieta\
transition, can be easily estimated as was done in Ref.~\cite{Kud}.

The propagation of the physical states is described by the eigenvalues
of $\mathbf{G}$, i. e.
\be
G_{a_0}=2/(g_{11}+g_{22}+r) \ \mbox{and} \ G_{f_0}=2/(g_{11}+g_{22}-r) \ ,
\label{physprop}
\ee
where $r = \sqrt{(g_{11}-g_{22})^2+4g_{12}^2}$.
The physical masses of the scalar mesons $a_0$ and $f_0$ are given
by the real part of the zeros of Eq.~(\ref{physprop}).

As usual the full propagator $\mathbf{G}$ can be expressed in terms of 
the physical propagators $G_{a_0}$ and $G_{f_0}$ plus a parameter for 
the mixing, which can be identified as
$
\delta = 2g_{12}/(g_{11}-g_{22}) \ .
$
In the case of $\rho$--$\omega$ mixing or $\pi$--$\eta$ mixing
$\delta$ is guaranteed to be small due to a large denominator $g_{11}-g_{22}$.
In the former case there is a large difference in the widths whereas
in the latter the masses of the mixing particles are rather different.
However, the experimental evidence that is available for the 
masses and widths of the scalar mesons $a_0$ and $f_0$ suggests that 
here both these quantities might be very similar \cite{PDG}.
Thus, even $g_{11}=g_{22}$ is not excluded.
Also, we want to emphazise that
$\delta$ can not be interpreted as a mixing angle because
it is necessarily complex valued. 

For convenience we write the primary production amplitude 
$\cal M$ as a vector in isospin space,
\begin{eqnarray}
{\mathbf{M}} =  
\left( \begin{array}{c}
{\cal M}_1 \\  {\cal M}_0 
\end{array} \right) \ ,
\nonumber 
\end{eqnarray}
and the final production vertex $W_X$ as a matrix
\begin{eqnarray}
{\mathbf{W}} = 
\left( \begin{array}{cc}
W_1^{(\pi \eta)} &  W_1^{(\pi \pi)} \\  W_0^{(\pi \eta)} & W_0^{(\pi \pi)}
\end{array} \right) \ .
\nonumber 
\end{eqnarray}
Note that here $W_1^{(\pi \eta)}$ is the vertex
for the coupling of the $\pi \eta$ system to an isospin-one particle as 
given by an elementary Lagrangian, whereas
$W_0^{(\pi \eta)}$ is the vertex where the $\pi \eta$ system mixes
first into the $\pi \pi$ system which then couples to an isospin-zero 
particle. We thus find
for the complete production amplitude of the final states $X$ 
= $(\eta \pi)$ or $(\pi\pi)$ 
(c.f. Eq. (\ref{adef}) for the one channel situation)

\begin{eqnarray}
{\mathbf{A}} = {\mathbf{W}}^\dagger{\mathbf{G}}{\mathbf{M}} 
= {\mathbf{W}}^\dagger \frac{1}{g_{11}g_{22}-g_{12}^2}
\left( \begin{array}{cc}
g_{22} & -g_{12} \\  -g_{12} & g_{11}
\end{array} \right) {\mathbf{M}} =  
{\mathbf{W}}^\dagger G_{a_0} G_{f_0} \left( \begin{array}{cc}
g_{22} & -g_{12} \\  -g_{12} & g_{11}
\end{array} \right) {\mathbf{M}} \ .
\nonumber 
\end{eqnarray}

Basically all investigations on the \mix so far don't use
the propagators of the physical states, i.e. they don't
use the eigenvalues of $\mathbf{G}$. Rather they assume that
the $a_0$ and $f_0$ mesons still propagate as proper isospin
states. 
This leads to expressions that are very easy to interpret and
which are illustrated in Fig. 2 for the case of $X=(\pi \eta)$.
Within the formalism developed above this assumption 
corresponds to the approximation 
\begin{equation}
G_{f_0}G_{a_0}=\frac{1}{g_{11}g_{22}-g_{12}^2} \simeq 
\frac{1}{g_{11}g_{22}} = \tilde G_{f_0} \tilde G_{a_0},
\label{gapprox}
\end{equation}
where $\tilde G_{f_0}$ and $\tilde G_{a_0}$ are now propagators of these
mesons in the isospin $0$ and $1$ states, respectively. 
Indeed, the results presented in
Ref.~\cite{Kud} suggest that
$$
\frac{g_{12}^2}{g_{11}g_{22}}<10^{-2} \ , 
$$
i.e. that the approximation (\ref{gapprox}) is not unreasonable. 

Since the main focus of the paper is the $\pi \eta$ channel it is convenient
to introduce a mixing parameter $\xi$ by
\be
\xi(m) = \Sigma_{01}(m)\, \tilde G_{f_0}(m^2), 
\label{52}
\ee
where $\Sigma_{01}$ was defined in Eq.~(\ref{smix}) and is related to the
\mix transition amplitude. With this definition we get for the
transition amplitude for the reaction $pn \to d\pi \eta$
\begin{equation}
{\cal A}_{(\pi \eta)} = W^{(\pi \eta)}_1\tilde G_{a_0}({\cal M}_1
+\xi (m){\cal M}_0)+ W^{(\pi \eta)}_0\tilde G_{f_0}{\cal M}_0 
+ W^{(\pi \eta)}_0\xi (m) \tilde G_{a_0} {\cal M}_1 \ .
\label{mmm}
\end{equation}
The last term involves mixing in $W_0^{(\pi\eta)}$ as well as in $\xi (m)$ 
and is therefore a second order correction. 
Close inspection reveals that in the region $m\simeq 2 m_K$ the
third term in Eq.~(\ref{mmm}) should be suppressed by a factor
$$
\frac{ g_{f_0\pi\pi} |\lambda_{\pi^0\eta}| m_{a_0} \Gamma_{a_0}}
{ g_{a_0\pi\eta} \Sigma_{01} (m^2_{\eta}-m^2_{\pi})} \approx 0.12 
$$
compared to the second term. In this estimation 
we used the values $\Sigma_{01}\approx 5000$~MeV$^2$  determined from 
$K\bar K$ decay loops (see Fig.~2 in Ref.~\cite{Kud}) at
$m\simeq 2m_K$ and the $\pi\eta$ mixing amplitude 
$\lambda_{\pi^0\eta}\simeq -5000$~MeV$^2$
(see Ref.~\cite{Kud} and references therein). We also use the
masses and widths $m_{a_0}=m_{f_0}=980$~MeV$/c^2$ and
$\Gamma_{a_0}=\Gamma_{f_0}=50$~MeV$/c^2$ and coupling constants $g_{a_0\pi\eta}$
and $g_{f_0\pi\pi}$ that are determined from the equations
$\Gamma_{a_0}= g^2_{a_0\pi\eta} q_{\pi\eta}/(8\pi m^2_{a_0})$ and
$\Gamma_{f_0}= 3 g^2_{f_0\pi\pi} q_{\pi\pi}/(16\pi m^2_{f_0})$,
where $q_{\pi\eta}$ and $q_{\pi\pi}$ are corresponding relative momenta.
Consequently, both terms -- the third and the last in Eq.~(\ref{mmm})
-- will be neglected in what follows.

\section{Structure of the primary production amplitude}

As we argued in the former section one can safely assume
the primary production amplitudes $\cal M_I$ as isospin conserving.
In this section we demonstrate how to construct the effective
interaction relevant for the transitions $pn\to d+(scalar \ meson)$ in
the close-to-threshold regime.

As pointed out in Refs.~\cite{Gri,Kud}, if the scalar meson has isospin 1
it can only be produced in a $p$ wave. Since the initial $pn$ system
has to be in an isotriplet state and has to have odd parity, the Pauli
principle requires that it is also in a spin-triplet state.
Thus, the effective transition operator for the isovector final state
has to be linear in $\bfk$ and has to have an odd power of $\bfp$
(hereafter $\bfk$ and $\bfp$ denote the final and initial relative
three-momenta in the considered reaction, respectively).
It also has to be linear in both the spin
$\bfs :=\phi^T_1\sigma_2\bfsigma\phi_2$ of the initial nucleons pair
and the polarization vector $\bfeps$ of the outgoing deuteron. These
constraints lead to the reaction amplitude of the following type
\be
{\cal M}_1 = a\, (\bfp\cdot\bfs)\, (\bfk\cdot\bfeps^*)
    + b\, (\bfp\cdot\bfk)\, (\bfs\cdot\bfeps^*)
    + c\, (\bfk\cdot\bfs)\, (\bfp\cdot\bfeps^*)
    + d\, (\bfp\cdot\bfs)\, (\bfp\cdot\bfeps^*)\, (\bfk\cdot\bfp)\, .
\label{8}
\ee
The coefficients $a$, $b$, $c$ and $d$ are independent scalar amplitudes.
They may depend on the total CM energy and are necessarily complex since
they contain the initial state interaction. For the isospin 1 amplitude
relevant here the phase induced by the two nucleon unitarity cuts
can be related to the $NN$ phase shifts, as pointed out in
Ref.~\cite{kanzoandi}. However, for the isoscalar initial state
(required for the production of the $f_0$ meson, cf. below) there is no
phase-shift analysis available at the relevant energies due to a lack of
high energy $pn$ scattering experiments \cite{Arndt}. Therefore, in this
paper we will not consider the effects induced by the initial state
interaction. Let us mention, however, that their effects do not influence
the qualitative aspects discussed here. Note, that
the scalar amplitudes $a$, $b$, $c$ and $d$ are expected to have quite
smooth energy dependence and, accordingly, can be considered as constant
in the near threshold region.

In Ref. \cite{Oset} the transition amplitude was given in a partial wave
decomposed form. Obviously the two descriptions are equivalent, however,
we regard Eq. (\ref{8}) as more convenient for the construction of polarization
observables.

The amplitude relevant for the near-threshold production of an
isoscalar scalar meson can be constructed along the same lines and is given by
\be
{\cal M}_0 = f\, (\bfs\cdot\bfeps^*)
    + g\, (\bfp\cdot\bfs)\, (\bfp\cdot\bfeps^*)\, .
\label{81}\ee
Similar expressions for the amplitudes~(\ref{8}) and~(\ref{81})
can be found in Ref.~\cite{Grish,Kondr}. Note, however, that the terms
proportional to $g$ and $d$ were omitted in that work.

It is obvious from Eqs.~(\ref{8}) and (\ref{81}) that 
${\cal M}_1 \propto k$ and ${\cal M}_0 \propto const.$
near the threshold. For the isospin conserving case we therefore get
$$
\frac{d\sigma}{dm^2} (pn\to da_0) \propto Q^{3/2} \ \mbox{and}
 \ \frac{d\sigma}{dm^2} (pn\to df_0) \propto Q^{1/2} \ ,
$$
where $Q(m) = \sqrt{s}-M_d-m$ is the excess energy. Thus, if we study the
production of an isospin 1 final state, namely $\pi^0\eta$, the mixing
with the isoscalar ($f_0$) state will be kinematically enhanced \cite{Kud}.

\section{Study of observables}

\subsection{Invariant mass spectrum}
Naturally the simplest observable is just the invariant mass
spectrum of the reaction $pn\to dX$. As long as we look at
energies/values of $m^2$ such that $Q(m^2)$ is small, the spectrum
should show a resonant peak $\sim {\sum\limits_{X}}\rho_X$ 
reflecting the $m^2$ dependence of the $f_0$ propagator, 
cf. Eq. (\ref{rhoeig}).
It is possible to deduce the branching ratios of a particular resonance
by a fit of the Flatt\'e distribution to the mass spectrum. This was
demonstrated, e.g., for the case of the $a_0^+$ in Ref.~\cite{BNL}.
Since so far little is known about the $f_0$ this very easy to measure
observable is very interesting.
In addition, the number of all events in this peak is proportional to
$$
|{\cal M}_0|^2= 3 |f|^2 + 2\,{\rm Re}(fg^*) p^2 + |g|^2 p^4\, ,
$$
c.f. Eqs. (\ref{rhoeig}), (\ref{csform}) and (\ref{81}).
Note that the knowledge of the amplitudes $f$ and $g$ is important,
if we want to deduce quantitative informations about the $f_0/a_0$ mixing
amplitude from the reaction $pn\to dX$.

Definitely the most interesting observables are those concerning
the channel $pn\to d\pi \eta$,
since here the isospin conserving amplitude enters
in a $p$ wave whereas the isoscalar amplitude, that can be mixed in via
isospin violation, enters in an $s$ wave. This is the kinematical
enhancement mentioned at the end of the last section.

Let us discuss the mass distribution $d\sigma/dm_{\pi^0\eta}$ for the
$\pi^0\eta$ system in the reaction $pn\to d\pi^0\eta$. Note that near
the threshold of the reaction $pn\to d a^0_0$ this mass distribution
should be sensitive to the magnitude of \mix mixing as well as to
the mixing mechanism.

Consider first the limiting case of isospin conservation, so that the
\mix mixing is absent. In this case the mass spectrum
of the $\pi^0\eta$
system should coincide with the spectrum of the $\pi^+\eta$ system from
the reaction $pp\to d a^+_0\,$ -- apart from an overall (isospin)
factor of 0.5 \cite{Grish}. The amplitude of the reaction
$pn\to d\pi^0\eta$ is then given by the diagram of Fig.~2a.
In this case the invariant
$\pi^0\eta$-mass distribution is determined by the phase space
and $\rho_{(\pi \eta)}$, which was defined in Eq. (\ref{rho}), and is
dominated by the $a_0$ propagator (c.f.
Eq.~(\ref{prop})). The $\pi^0\eta$-mass distribution obtained under these 
asumptions, is shown in Fig.~3 by the dotted line. The calculations
are done at $T_p=2645$~MeV$/c$ where experimental results can be
expected from the ANKE collaboration soon \cite{Kondr}.
The parameters used here are: $M_{a_0}=980$ MeV$/c^2$,
$\Gamma_{a_0}=50$ MeV$/c^2$. The mass-dependent width is described by a 
Flatt\'e-like form $\Gamma_{a_0}(m)=\Gamma_{a_0} +\Gamma_{K\bar K}(m)$,
which takes into account the $K\bar K$ decay channel
(see Eq.~(8) in Ref.\cite{Kud}).
As we discussed in the previous
Sections, the isospin conserving production of the $a^0_0$ meson
can take place only in a $p$ wave and hence is suppressed near threshold.

If isospin is not conserved and \mix mixing is present, then the
$\pi \eta$ system can be
produced also in an $s$ wave through the $f_0$. The
corresponding diagram is shown in Fig.~2b. 
For this mechanism the $\pi^0\eta$ mass 
spectrum is determined not only by the $a^0_0$ propagator, 
the partial wave of production and the phase space but also
by the $f_0$ propagator and by the mass dependence of the nondiagonal
self-energy matrix element $\Sigma_{01} (m)$ 
(cf. Eqs.~(\ref{smix}) and (\ref{52})): 
\begin{eqnarray}
\nonumber 
\xi(m) = \Sigma_{01}(m)\, \tilde G_{f_0}(m^2) \approx 
\frac{\Sigma_{01}(m)}{m^2-M^2_{f_0} +i M_{f_0} \Gamma_{f_0}(m^2)}\, .
\end{eqnarray}
For illustrative purposes we will use 
 $M_{f_0}=980$ MeV$/c^2$ and the same Flatt\'e-like width 
that was used for the $a_0$ meson above, i.e.
$\Gamma_{f_0}(m^2)=\Gamma_{a_0}(m^2)$.

Of course, now the invariant mass plot becomes sensitive to the specific
mixing mechanism. For example, if we assume that
the leading contribution to $\Sigma_{01} (m)$ comes from the direct
\mix transition mechanism, i. e. the term $\alpha$ in Eq.~(\ref{smix}),
then there should be no drastic dependence on the invariant mass
$m$ of the $\pi^0\eta$ system.
The $\pi^0\eta$-mass spectrum, corresponding to the mechanism of Fig.~2b
and with $\Sigma_{01} (m)=const$, is shown in Fig.~3 by the dashed line.
(All the distributions are normalized to 1 at the maximal values!) 
On the other hand,
if $\Sigma_{01} (m)$ is determined mainly by the $K\bar K$ loop  (the
term $\Pi_{K^+ K^-}$$\,-\,$$\Pi_{\bar K^0 K^0}$ in Eq.~(\ref{smix})),
it should be strongly enhanced near the $K\bar K$ threshold (see, e.g.,
Refs.~\cite{Acha,Krepl,Kud}). 
The corresponding $\pi^0\eta$-mass distribution
is shown in Fig.~3 by the solid line.

Thus, we conclude that the $\pi^0\eta$ mass spectrum is rather sensitive
to the \mix mixing mechanism. That is why the study of
$d\sigma/dm_{\pi^0\eta}$ in the reaction $pn\to d\pi^0\eta$ should
allow to shed light on the nature of the lightest scalar mesons
$a_0(980)$ and $f_0(980)$.

\subsection{Analyzing power for the process $pn\to d\pi\eta$}
\subsubsection{Near threshold}

In our approximation the total amplitude $\cal M$ of the resonant
$\pi \eta$--production
with \mix mixing effects taken into account may be written as
(c.f. Eq. (\ref{mmm}))
\be
{\cal M} = {\cal M}_1 + \xi {\cal M}_0\, ,
\label{51}
\ee
where the amplitudes ${\cal M}_1$ and ${\cal M}_0$ are given by the
Eqs.~(\ref{8}) and (\ref{81}) and the mixing parameter $\xi$ was
defined in Eq. (\ref{52}). 
 
Let us introduce the polarization of one of the nucleons $\bfzeta$
through $\bfzeta=\phi^+_1\bfsigma\phi_1\,$ or
$\,\phi_1\phi^+_1 =(1+\bfzeta\cdot\bfsigma)/2$.
The matrix element~(\ref{51}) squared and averaged (summed)
over the polarizations of the initial neutron (final deuteron) is then
given by 
$$
\overline {|{\cal M}|^2}=\frac{1}{2}
\left[\left( |a|^2 +|c|^2\right)p^2 k^2 + 3 |B|^2 + |D|^2 p^4 \right]
+(\bfp\cdot\bfk)\, {\rm Re}H
$$
\be
+ p^2\, {\rm Re}\left[a^* D\, (\bfp\cdot\bfk) + B^* D\right]
+ (\bfzeta\cdot[\bfk\times\bfp\,])\, {\rm Im}H \, ,
\label{54}
\ee
where
$$
B=b\, (\bfp\cdot\bfk) +f\,\xi\, ,
\phantom{xx}
D=d\, (\bfp\cdot\bfk) +g\,\xi\, ,
\phantom{xx}
H= a^* B + a^* c\, (\bfp\cdot\bfk) + B^* c +D^* c\, p^2\, .
$$
Let the polarization vector $\bfzeta$ be directed along the $x$-axis
($\bfzeta\perp\bfp$). Then, in terms of the angles $\theta$ and $\varphi$,
we have $\bfp\cdot\bfk=pk\cos\theta\,$ and
$\bfzeta\cdot[\bfk\times\bfp\,]=\zeta\, p k \sin\theta\,\sin\varphi\,$
($\zeta=|\bfzeta|$). Using those expressions Eq.~(\ref{54}) may be 
written as
\vspace{1mm}
\be
\overline {|{\cal M}(\theta,\varphi)|^2}=
C_0 + C_1 \cos\theta +C_2 \cos^2\theta
+ \,\zeta\, \sin\theta\,\sin\varphi\, (D_0 + D_1\cos\theta)\, ,
\label{55}\ee
where
$$
C_0 = \frac{1}{2} \left( |a|^2 +|c|^2\right) p^2 k^2
    + \left( |f|^2 + \frac{1}{2} |f+p^2 g|^2 \right)\, |\xi|^2\, ,
$$ $$
C_1 = pk\, {\rm Re} \left(\left[ (a+b+c+p^2 d)^* (f+p^2 g)
    + 2b^* f \right]\,\xi \right)\, ,
$$
\be
C_2 = p^2 k^2\, \left[ |b|^2 +\frac{1}{2} |b+p^2 d|^2
    + {\rm Re} \left( a^* c + (a+c)^* (b+p^2 d) \right) \right]\, ,
\label{56}\ee
$$
D_0 =p k\, {\rm Im}\left( \left[ a^* f -c^* (f+p^2 g)\right]
\xi \right)\, ,
$$ $$
D_1 =p^2 k^2\, {\rm Im} (a^* b + a^* c + b^* c + p^2 d^{\,*} c)\, .
$$
\vspace{1mm}
We can now generalize the definition of the asymmetry $A$
given in Ref.~\cite{Kud} to the polarized situation. 
Using the short hand notation
\be
\sigma (m;\theta, \varphi) \equiv \frac{d^2\sigma}
{dm^2d\Omega } (\theta, \varphi)\, .
\label{57}\ee
we now define the angular--asymmetry parameter $A$ through
\begin{eqnarray}
\nonumber 
A(m;\theta,\varphi)&=&
\frac{\sigma (m;\theta, \varphi) -\sigma (m;\pi-\theta, \varphi+\pi)}
{\sigma (m;\theta, \varphi) +\sigma (m;\pi-\theta, \varphi+\pi)}\, \\
&=&\frac{C_1 \cos\theta\,
+\,\zeta\, D_0 \sin\theta\,\sin\varphi }
{C_0 + C_2 \cos^2\theta
+ \,\zeta\, D_1 \cos\theta\,\sin\theta\,\sin\varphi}\, .
\label{58}
\end{eqnarray}
It follows from Eqs.~(\ref{56}) and~(\ref{58}) that $C_1=D_0=0$ for $\xi=0$
and thus 
\be
A(m;\theta, \varphi)_{\xi=0}=0\,  
\label{591}
\ee
when isospin is conserved. 

Let us discuss some specific features of the expression~(\ref{58}) in
the following. First note that there are two different terms. 
The first term does not depend on the 
polarization $\zeta$ of the proton beam and on the angle $\varphi$. 
Therefore, Eq.~(\ref{58}) implies that 
\be
A(m;\theta=0^0, \varphi)=\frac{C_1}{C_0+C_2} \ .
\label{59}
\ee
This particular result for the 
asymmetry $A$ was derived earlier in Ref.~\cite{Kud}
but not in terms of the amplitudes $a$, $b$, $c$, $d$, $f$ and $g$.
The discussion in Refs.~\cite{Grish,Kondr}, on the other hand, 
takes into account only the amplitudes $a$, $b$, $c$, and $f$,
but not the amplitudes $d$ and $g$.

At $\theta=90^0$ and for $\zeta\ne 0$ we get
\be
A(m;\theta=90^0, \varphi)=\zeta\,\frac{D_0}{C_0}\,\sin\varphi\, =
\frac{2\zeta\,p k\, {\rm Im}\left( \left[ a^* f -c^* (f+p^2 g)\right]\,
\xi \right) \sin\varphi} {\left( |a|^2 +|c|^2\right) p^2 k^2
    + \left( 2 |f|^2 + |f+p^2 g|^2 \right)\, |\xi|^2}\, .
\label{510}
\ee

Obviously
the isospin-breaking effect for different angles $\theta$ and
$\varphi$ depends on different combinations of the basic amplitudes.
Thus, experimental information on the asymmetry obtained with polarized 
beam could allow to deduce additional constraints on the amplitudes $a$, 
$b$, $c$, $d$, $f$, and $g$ of the $a_0$- and $f_0$ production reactions.
With regard to that let us mention that 
the model of $a_0$-meson production discussed in
Ref.~\cite{Kud}, which was based on the impulse approximation, leads to
$A(m;\theta=90^0, \varphi)\equiv 0$ -- because in that model $a = c = 0$.

Throughout this paper we make the assumption that isospin breaking
takes place only in the propagation of the scalar mesons. This implies
that the mixing strength is the same for all partial waves.
On the other hand, if a significant mixing takes place also in the 
initial $NN$ interaction or in the primary production amplitude 
${\cal M}_1$, there is no reason to expect this mixing as being 
partial-wave independent. Thus the study of polarization observables allows 
to examine this basic model assumption in a clean way.

In addition it will be very interesting
to study the $m$ dependence of $A(m;\theta,\varphi)$ in detail. 
By construction $A$ projects on the isospin breaking pieces of
the amplitude. Therefore, the $m$ dependence of $A$ gives direct access
to the $m$ dependence of $\Sigma_{01}(m)$ and thus to the mixing
mechanism.

\subsubsection{Higher energies}

As mentioned above the asymmetry $A(m;\theta, \varphi)$ (Eq.~(\ref{58}))
will become zero if \mix mixing is not included, i.e. for $\xi=0$. This
result is in line with the observation made in Ref.~\cite{Miller}
that the asymmetry $A(\theta)$ is identical to zero in the reaction 
$pn\to d\pi^0$ for the unpolarized case. 
Let us emphasize, however, that the vanishing of the asymmetry
$A(m;\theta, \varphi)$ in the limit $\xi=0$ for the polarized case is true
only near threshold, where the $a_0$ meson is produced in a $p$ wave and
the $f_0$ meson in an $s$ wave, see Eqs.~(\ref{8}) and (\ref{81}). 
The expressions~(\ref{8}) and (\ref{81}) do not take into account
contributions from higher partial waves, 
specifically they do not include, e.g., the $d$ wave production of the 
$a_0$ meson and the $p$ wave production of the $f_0$ meson. 
In the following we want to consider this question in more detail,
assuming that isospin is conserved. 

The amplitudes for $a_0$- and $f_0$ production in the
reaction~(\ref{1}) can be written in the most general form as
\be
{\cal M}_I =\phi^T_1\sigma_2\, (F_I +\bfg_I\cdot\bfsigma)\,\phi_2 \, ,
\label{511}
\ee
where $\phi_{1,2}$ are the spinors of the nucleons. The two terms 
$F_I$ and $\bfg_I$ correspond to $S_{NN}=0$ and $S_{NN}=1\,$,
respectively, where $S_{NN}$ is the total spin of the initial $NN$
system. The subscript $I$ denotes the isospin of the produced scalar
meson. The scalar and vector
functions $F_I$ and $\bfg_I$ depend on the vectors $\bfp$, $\bfk$ and
$\bfeps$. Let $L_{NN}$ and $L_{Md}$ be the angular momenta in the
initial $NN$ and in the final meson$+d$ systems, respectively. It follows
from the conservation of quantum numbers together with the required 
antisymmetry of the system with respect to the initial nucleons that 
the $F_1$ ($\bfg_1$) term in Eq.~(\ref{511}) should contain only the
contributions from even (odd) values of $L_{NN}$ and $L_{Md}$.
One can also see that the $F_0$ ($\bfg_0$) term should only contain the
contributions from odd (even) values of $L_{NN}$ and $L_{Md}$.
In the isovector ($I=1$) case the $s$ wave ($L_{NN}=L_{Md}=0$) is
forbidden, i.e. the contributions to the $F_1$ term start with a 
$d$ wave.
For the case where one of the nucleons is polarized, the expression 
for the squared matrix element~(\ref{511}) is given by the form 
\be
\overline {|{\cal M}_I|^2}=\frac{1}{2} \left[ |F_I|^2
+ \left( [ F_I \bfg^*_I + F^*_I \bfg_I ]\cdot\bfzeta\right)
+ (\bfg^*_I \cdot \bfg_I)
+ i \left(\bfzeta\cdot [\bfg_I \times \bfg^*_I ]\right)\right]\, ,
\label{512}
\ee
where $\bfzeta$ is the proton polarization vector. 
The second term in the
expression~(\ref{512}) corresponds to the interference between the
$F_I$ and $\bfg_I$ amplitudes. Thus, this term is an
odd function of the final momentum $\bfk$. 
All other terms in Eq.~(\ref{512}) are even functions of 
$\bfk$. We see that the second term is responsible for the angular
asymmetry in the reaction~(\ref{1}) even if isospin is conserved.
 It vanishes in the case of
unpolarized protons, i.e. $A(m;\theta, \varphi)\equiv 0$ for $\bfzeta=0$.

Note that the amplitudes~(\ref{8}) and (\ref{81}) are special cases of
the general form Eq.~(\ref{511}) with $F_I\equiv 0$. 
Also, setting $F_I\equiv 0$ in 
Eq.~(\ref{512}) we always get $A(m;\theta,\varphi)\equiv 0$ if isospin is
conserved. The amplitudes~(\ref{8}) and (\ref{81}) correspond
to the $\bfg_I$ terms in Eq.~(\ref{512}) in the lowest-order
approximation with respect to the final momentum $\bfk$, i.e. keeping
only contributions that are at most linear in $\bfk$.
The next-order term with respect to $\bfk$ in the amplitude for
$a_0$-meson production is the $d$-wave contribution. This term (of the
order $\sim k^2$) contributes to $F_1$ in Eq.~(\ref{511}) and
modifies the amplitude ${\cal M}_1$ given in Eq. (\ref{8}), i.e.
\be
{\cal M}_1 \to {\cal M}_1 + F_1 \,\, \phi^T_1\sigma_2\,\phi_2 \, ,
\phantom{xxx}
F_1 =e\, (\bfeps^*\cdot [\bfp\times\bfk]) (\bfp\cdot\bfk)
\label{513}
\ee
where $e$ is a scalar amplitude. It is clear that the asymmetry
$A(m;\theta, \varphi)$~(\ref{58}), calculated with the
amplitude~(\ref{513}), will get non-zero contributions $\sim k^3$ for
small $k$ due to the interference between $p$ and $d$ waves.
It follows from Eq.~(\ref{513}) that $F_1=0$ at $\sin\theta=0\,$ or
$\cos\theta=0\,$. 
In general, the $F_1$ term contains all contributions with even values 
$L_{NN}, L_{Md} =2,4,6,...\,$. Thus, each contribution should include 
the factor $\bfeps^*\cdot [\bfp\times\bfk]$
as well as the factor $\bfp\cdot\bfk$. Therefore, even in the
general case $F_1=0$ if $\sin\theta=0\,$ or $\cos\theta=0\,$. 
Accordingly, for the case of conserved isospin we get 
\be
A(m;\theta=0,\varphi)\equiv 0
\label{5131}
\ee
at any $k$ -- and not only near threshold (compare this result with 
that from Eq.~(\ref{591})). This result (\ref{5131}) confirms a general
statement made in Ref.~\cite{Miller} on the absence of a forward--backward
asymmetry for the reaction $pn\to d X^0$ in the limit of isospin
conservation.

It is remarkable that in addition to the condition~(\ref{591}) for the
case of polarized proton beam we get also the nontrivial result that
\be
A(m;\theta=90^0,\varphi)\equiv 0
\label{514}
\ee
if isospin is conserved. 
Therefore, any deviation of $A(m;\theta=90^0,\varphi)$ from zero 
is a direct indication for \mix mixing. Note that
this statement does not depend on the number of partial waves
taken into account, i.e. it is valid for any excess energy and not
only near threshold. Consequently, a
measurement of the asymmetry at $\theta=90^0$
should provide us with evidence on the \mix mixing amplitude. 
We want to remark that this information is to be considered as 
complementary to the one that can be obtained with unpolarized beam.
It should be mentioned that the vanishing of the asymmetry at
$\theta=90^0$ for the reaction $pn\to d a^0_0\,$ does not follow from
the theorem formulated in Ref.~\cite{Miller} and is new.

Finally, let us discuss the situation at $\theta\ne 0^0$ and 
$\theta\ne 90^0$.
If the amplitude for $a^0_0$-meson production is taken
of the form~(\ref{51}), i.e. isospin-breaking effects are included, 
then the leading-order contribution to the asymmetry
$A(m;\theta,\varphi)$ are of the order $\sim \xi k$ 
(see Eqs.~(\ref{56}) and (\ref{58})). The
next-to-leading terms in the amplitude ${\cal M}_1$
(see Eq.~(\ref{513})) give rise to contributions of the order $\sim k^3$
to the asymmetry from isospin-conserving terms. Thus, when studying
isospin-breaking effects over a wide region of $\theta$, one should
consider $a^0_0$-meson production in a rather narrow region of 
relative momenta $k$, i.e. at small excess energies,
in order to suppress contributions to $A(m;\theta,\varphi)$ from
isospin-conserving amplitudes with higher angular momenta.

\section{Summary}

Let us summarize the main results of this paper.
The most sensitive observables for examining the \mix mixing amplitude are:

\begin{itemize}
\item[{\it i})]
the angular asymmetry $A(m;\theta, \varphi)$, as defined by Eq.~(\ref{58})
of this paper 
\item[{\it ii})]
the distribution of the effective mass of the $\pi^0\eta$ system near the
$a^0_0$ threshold 
\end{itemize}
With regard to the 
angular asymmetry $A(m;\theta, \varphi)$ for the channel
$pn\to d\pi^0\eta$, it is expected to be rather large, i.e. in the order of
$\sim 10^{-1}$. For the case of an unpolarized beam the forward-backward
asymmetry was estimated earlier in Ref.~\cite{Kud}. The study of the
asymmetry $A(m;\theta, \varphi)$ with perpendicular polarized proton
beam should shed additional light on the mechanism of $a_0$ production
in the reaction $pn\to d a^0_0$. As we have shown the various basic
amplitudes for this reaction give different contributions to
$A(m;\theta, \varphi)$ at different angles.

Since by constuction $A$ projects on the isospin breaking pieces of
the amplitude, the $m$ dependence of $A$ gives direct access
to the $m$ dependence of $\Sigma_{01}(m)$ and thus to the mixing
mechanism.

Note, in experiments with polarized beams
 isospin-breaking effects give contributions $\sim \xi k$ to
the asymmetry $A(m;\theta, \varphi)$ in leading-order 
with respect to the final momentum $\bfk$. Isospin-conserving amplitudes
give contributions $\sim k^3$ due to $d$-wave terms. Thus when
studying isospin-breaking effects over a wider range of $\theta$ and
$\varphi$, one should consider $a^0_0$-meson production
at small momenta $k$, or small excess energies. Accordingly, 
when studying the reaction $pn\to d\pi^0\eta$ near the
threshold of $a_0$ production, one should limit the invariant mass $m$ 
of the $\pi^0\eta$ system in a narrow region near threshold.
Note that the contribution of the $d$-wave to the reaction $pn\to d\,a_0$
may be monitored by a simultaneous measurement of the differential
cross section.

If isospin is conserved, then asymmetry effects should be absent
at $\theta=0^0$ or $\theta=90^0$, even in the polarized case.
The angle $\theta=0^0$ is not conclusive 
for the reaction with polarized beam, since all polarization
effects are proportional to $\zeta\sin\theta\cos\varphi$ and vanish at
$\sin\theta=0$. Thus, the asymmetry for $\theta=0^0$ coincides with 
the one for unpolarized nucleons.
The case $\theta=90^0$ is much more interesting
when studying the isospin-breaking effects in the reaction with
polarized protons, since polarization effects are maximal at
this angle. Anyway, in either case ($\theta=0$ and $\theta=90^0$) a 
non-vanishing 
asymmetry $A(m;\theta, \varphi)$ can come only from isospin-breaking
effects.

Very important informations on \mix mixing can be also expected from 
a study of the mass distribution $d\sigma/dm_{\pi^0\eta}$ for the
$\pi^0\eta$ system in the reaction $pn\to d\pi^0\eta$. If the 
isospin is conserved then 
the mass spectrum of the $\pi^0\eta$ system from this reaction
should coincide with the one for the $\pi^+\eta$ system from the reaction
$pp\to d a^+_0\,$. Differences in the mass distributions of the 
$\pi^0\eta$ and the $\pi^+\eta$ systems
are a clear signal that the isospin is not conserved and that
\mix mixing is present. The mass distribution $d\sigma/dm_{\pi^0\eta}$ 
for the $\pi^0\eta$ system is also extremely sensitive to the mechanism
of \mix mixing. For example, in the case where \mix mixing is
dominated by the $K\bar K$ loop, the $\pi^0\eta$ mass
spectrum will be strongly enhanced near the $K\bar K$ threshold.

Thus, we conclude that the study of the reaction $pn\to d a^0_0$ is
extremely useful to understand the mechanism and the magnitude of the
\mix mixing amplitude. The knowledge of this mixing amplitude should
allow us to 
shed light on the nature and the quark content of the lightest 
scalar mesons $a_0(980)$ and $f_0(980)$.

Several interesting observables where not discussed in this
paper and are to be studied in a subsequent work. 
One example is the differential cross section
for $pn\to dK^+ K^-$.
Since this final state does not have definite isospin the differential 
cross section develops a forward--backward asymmetry at larger 
excess energies $Q$. 
Models predict that the kaon loops are an important source for
the mixing \cite{Krepl}. Therefore, the measurement of
a forward--backward asymmetry for the process $pn\to dK\bar K$ (the sum
of $pn\to dK^+ K^-$ and $pn\to dK^0 \bar K^0$) should help to pin down the
\mix mixing amplitude. Corresponding measurements of these 
decay channels should be feasible at the ANKE and TOF facilities, 
respectively, at the COSY accelerator in J\"ulich. 

Finally let us consider the reaction
$dd\to {^4}{\rm He}\,\pi^0\eta$. Evidently, here isospin is not
conserved. Near the $a_0$ threshold this reaction should
proceed only through the chain
$dd\to {^4}{\rm He}\,f_0\to {^4}{\rm He}\,a^0_0\to {^4}{\rm He}\,\pi^0\eta$.
Thus, the $\pi^0\eta$ mass spectrum of this reaction should 
be even more sensitive to the \mix mixing mechanism, than the one
for the reaction $pn\to d a^0_0$ \cite{Grish}.

\vfill
\section*{Acknowledgments}
The authors are thankful to L.A.~Kondratyuk and M.~B\"uscher
for stimulating discussions. 
This work was partly supported by the DFG-RFBI grant No. 02-02-04001
(436 RUS 113/652/1-1) and by the RFBR grant No. 00-15-96562.


\newpage

\begin{figure}
\caption{Graphical representation of the Dyson equation.
The thin solid line denotes the bare propagator whereas the
thick solid line stands for the dressed/physical propagator. The sum
appearing in the expression for the self energy is assumed to run over
all decay channels $X$.}
\label{dyson}
\end{figure}

\begin{figure}
\caption{Different contributions to $\pi \eta$ production in leading
order in the mixing. Here $f_0-a_0$ mixing is denoted by a circle and
$\pi-\eta$ mixing is denoted by a cross. 
Note that $f_0-a_0$ mixing can occur via $K\bar K$ loops as well
as via direct mixing, as explained in Sect. 3. The numbers in the
primary production vertices (denoted by a big circle) indicate the
isospin of the relevant amplitude.}
\label{pieta}
\end{figure}

\begin{figure}
\caption{Spectra of the $\pi^0\eta$ invariant mass for the reaction
$pn\to d\pi^0\eta$ at $T_p=2645$~MeV$/c$. The dotted line corresponds 
to a reaction mechanism where 
an $a_0$ meson is produced in a $p$-wave and then decays into the
$\pi^0\eta$ system. 
The dashed line corresponds to the production of an $f_0$ meson 
in an $s$-wave that mixes into an $a_0$ with $\Sigma_{10} =const$, cf. 
Eq. (\ref{smix}). 
The solid line shows results where $\Sigma_{10}$ is evaluated 
from the $K\bar K$ mixing mechanism. 
Note that all the distributions are normalized
to 1 at their maximal value.}
\label{fig3}
\end{figure}


\newpage
\vglue 1cm
\begin{figure}[tb]
\vspace{3cm}
\includegraphics{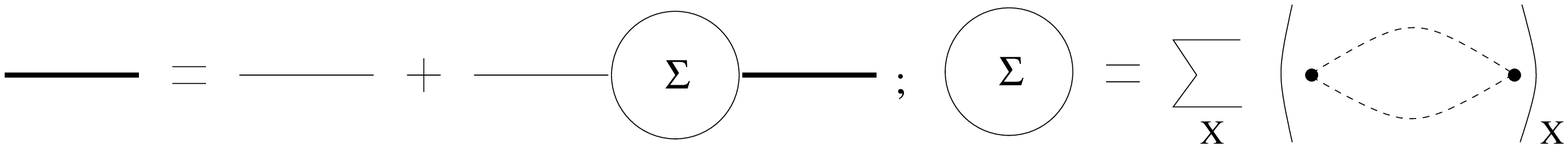}
\vspace{4mm}
\center{FIG. 1}
\end{figure}

\vskip 5cm 

\begin{figure}[tb]
\vspace{4cm}
\includegraphics{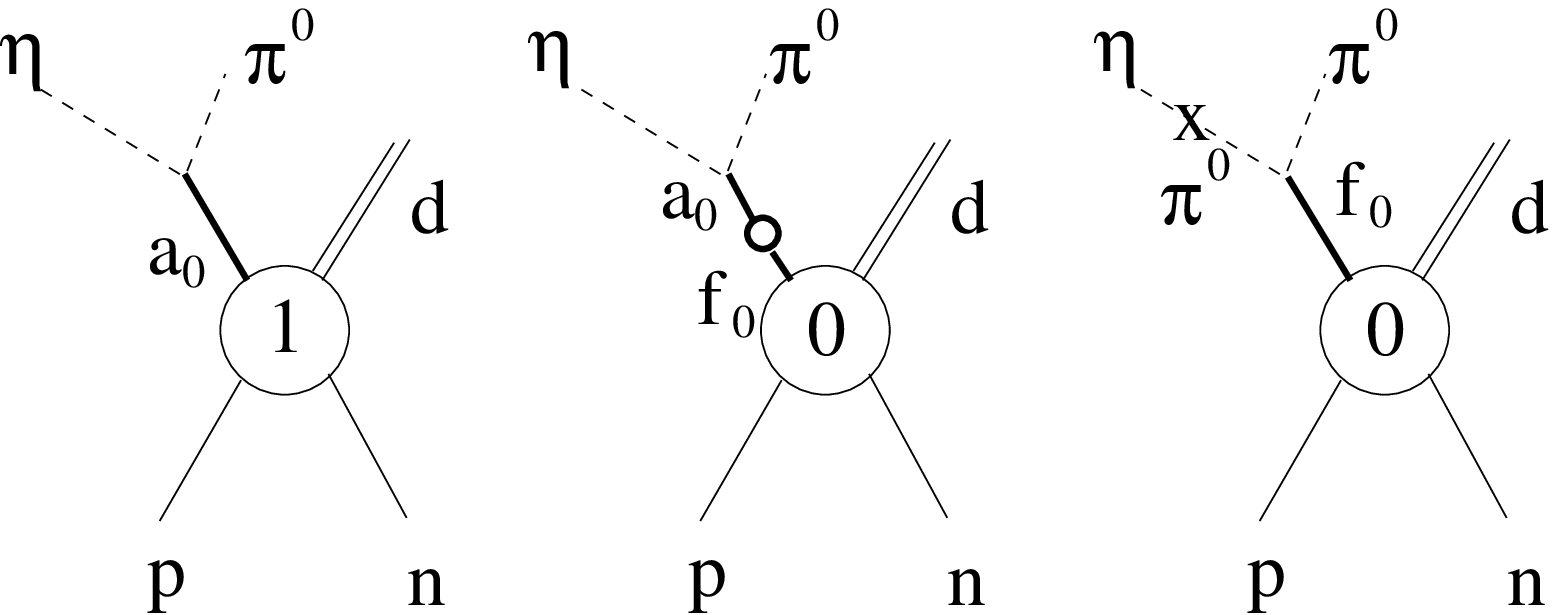}
\vspace{4mm}
\center{(a) \hskip 4cm (b) \hskip 4cm (c)}
\vspace{8mm}
\center{FIG. 2}
\end{figure}

\newpage

\vglue 1cm
\begin{figure}[t]
\vspace{13cm}
\includegraphics{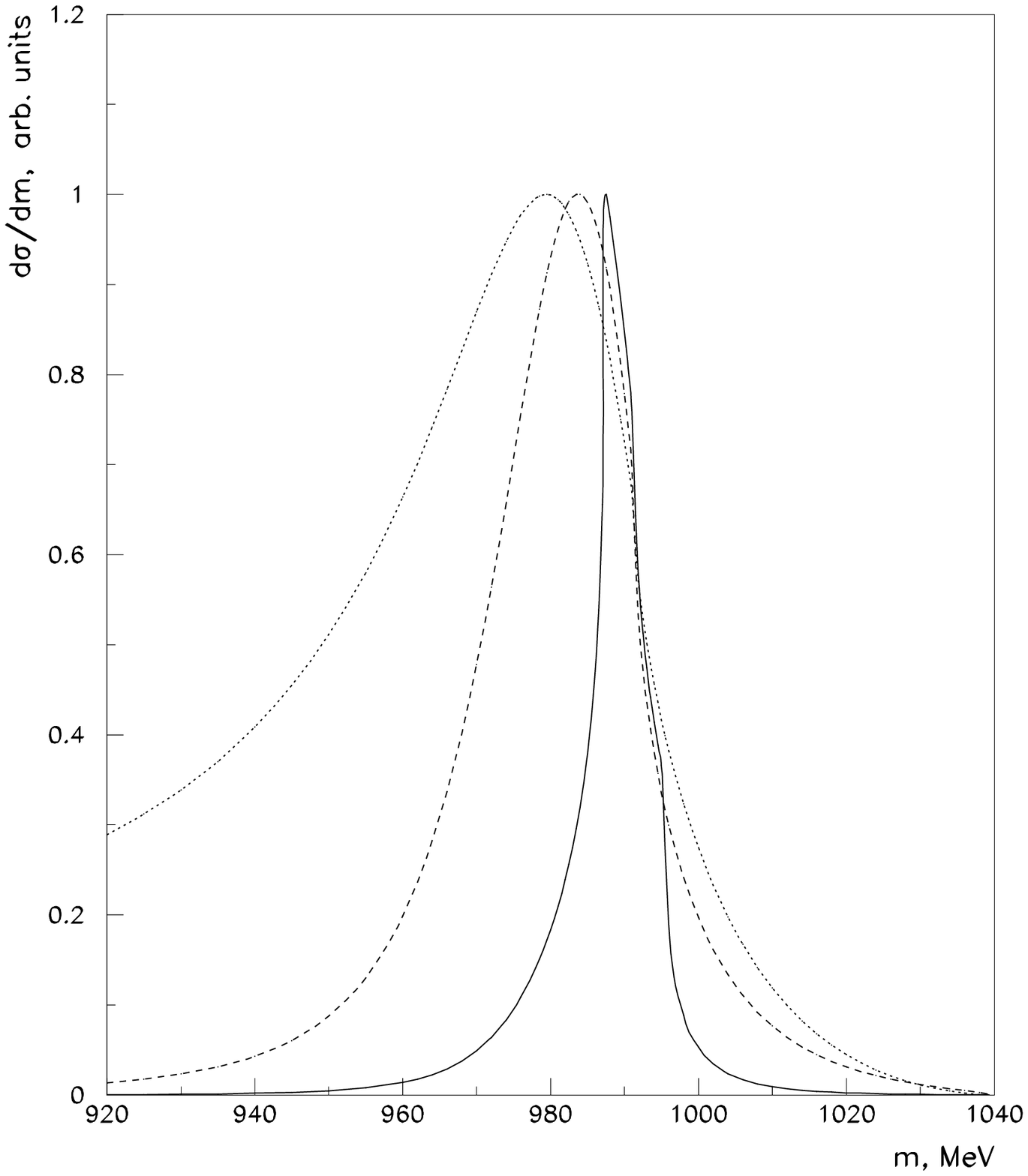}
\vspace{4mm}
\center{FIG. 3}
\end{figure}

\begin{thebibliography}{99}
%
\bibitem{Acha} N.N.~Achasov, S.A.~Devyanin, and G.N.~Shestakov,
                Phys. Lett. {\bf B88}, 367 (1979); Yad.Fiz. {\bf 33},
                1337 (1981) [Sov. J. Nucl. Phys. {\bf 33}, 715 (1981)].
%
\bibitem{Barnes} T. Barnes, Phys. Lett. {\bf B165}, 434 (1985).

\bibitem{Acha1} N.N.~Achasov and G.N.~Shestakov,
                Phys. Rev. D {\bf 56}, 212 (1997).

%
\bibitem{Krepl} O. Krehl, R. Rapp, and J. Speth,
                Phys. Lett. {\bf B390}, 23 (1997).
%
\bibitem{Kerb}  B.~Kerbikov and F.~Tabakin,
                Phys. Rev. C {\bf 62}, 064601 (2000).
%
\bibitem{Kud}  A.E.~Kudryavtsev and V.E.~Tarasov,
               JETP Lett. {\bf 72}, 401 (2000); {\tt nucl-th/0102053}.
%
\bibitem{Grish} V.Yu.~Grishina, L.A.~Kondratyuk, M.~B\"uscher, W.~Cassing, 
                and H. Str\"oher, Phys. Lett. {\bf B 521},
                217 (2001).
%
\bibitem{Close} F.E.~Close and A.~Kirk, 
                Phys. Lett. {\bf B489}, 24 (2000).
%
\bibitem{Barb} D.~Barberis, et al.,
               Phys. Lett. {\bf B488}, 225 (2000).
%
\bibitem{Kirk} A.~Kirk, Phys. Lett. {\bf B489}, 29 (2000).
%
\bibitem{Jans} G. Janssen, B. Pierce, K. Holinde, and J. Speth,
               Phys. Rev. D {\bf 52}, 2690 (1995).
%
\bibitem{Oller97} J.A. Oller and E. Oset,
               Nucl. Phys. {\bf A620}, 438 (1997);
               Nucl. Phys. {\bf A652}, 407 (1999).
%
\bibitem{weinberg} S.~Weinberg, 
Phys. Rev. {\bf 130}, 776 (1963);
Phys. Rev. {\bf 131}, 400 (1963).
%
\bibitem{Oset} E.~Oset, J.A.~Oller, and Ulf-G.~Mei{\ss}ner,
Eur. Phys. J. {\bf A12}, 435 (2001). 
%
\bibitem{flatte}  S.M.~Flatt\'e, Phys. Lett. {\bf 63}, 224 (1976).
%
\bibitem{Acha2} N.N.~Achasov, S.A.~Devyanin, and G.N.~Shestakov,
                Yad. Fiz. {\bf 32}, 1098 (1980)
                [Sov. J. Nucl. Phys. {\bf 32}, 566 (1980)].
%
\bibitem{Tipp} W.B. Tippens et al.,
               Phys. Rev. C {\bf 63}, 052001 (2001).
%
\bibitem{PDG} Particle Data Group: D.E. Groom et al.,
Eur. Phys. J. {\bf C15}, 1 (2000). 

\bibitem{Gri}   V.Yu.~Grishina, L.A.~Kondratyuk, E.L.~Bratkovskaya,
                M.~B\"uscher, and W.~Cassing, Eur. Phys. J. {\bf A9},
                277 (2000).
%
\bibitem{kanzoandi} C.~Hanhart and K.~Nakayama,
          Phys. Lett. {\bf B454}, 176 (1999).
%
\bibitem{Arndt} R.A. Arndt, I.I. Strakovsky, and R.L. Workman,
               Phys. Rev. C {\bf 62}, 034005 (2000).
%
\bibitem{Kondr} M.~B\"uscher et al., {\it Investigation of neutral
scalar mesons $a_0/f_0$ with ANKE, COSY proposal \# 97} (2001), 
http://ikpd15.ikp.kfa-juelich.de:8085/doc/Proposals.html
%
\bibitem{BNL} S.~Teige, et al.,
               Phys. Rev. D {\bf 59}, 012001 (1999).
%
\bibitem{Miller} G.A.~Miller, B.M.K.~Nefkens, and I.~\v Slaus,
                 Phys. Rep. {\bf 194}, 1 (1990).
\end{thebibliography}
\end{document}